\documentclass[12pt]{iopart}

\usepackage{graphicx}
\usepackage{color}

\begin{document}

\title{Wronskian method for bound states}

\author{Francisco M. Fern\'andez}

\address{INIFTA (UNLP, CONICET), Divisi\'on Qu\'imica Te\'orica,
Blvd. 113 S/N,  Sucursal 4, Casilla de Correo 16, 1900 La Plata,
Argentina}\ead{fernande@quimica.unlp.edu.ar}

\maketitle

\begin{abstract}
We propose a simple and straightforward method based on Wronskians for the
calculation of bound--state energies and wavefunctions of one--dimensional
quantum--mechanical problems. We explicitly discuss the asymptotic behavior
of the wavefunction and show that the allowed energies make the divergent
part vanish. As illustrative examples we consider an exactly solvable model
and the Gaussian potential well.
\end{abstract}

\section{Introduction}

Wronskians (or Wronskian determinants) are most useful for the analysis of
ordinary differential equations in general\cite{A69} and the Schr\"{o}dinger
equation in particular\cite{PC61}. Whitton and Connor\cite{WC73} applied a
remarkable Wronskian formalism to resonance tunnelling reactions and we have
recently proposed that a closely related method may be suitable for teaching
quantum scattering in advanced undergraduate and graduate courses in quantum
mechanics\cite{F11}.

The purpose of this paper is to show that the Wronskian method is also
useful for the study of bound states of one--dimensional quantum--mechanical
models. We believe that this variant of the approach is also suitable for
pedagogical purposes and enables a unified treatment of the discrete and
continuous spectra of simple quantum--mechanical models.

In section~\ref{sec:Schro} we convert the Schr\"{o}dinger equation
into a dimensionless eigenvalue equation and show how to apply the
Wronskian method to bound states. In section~\ref{sec:examples} we
illustrate the application of the approach by means of two
suitable examples. In section~\ref {sec:conclusions} we outline
the main results of the paper and draw conclusions. Finally, in an
Appendix we outline the main properties of Wronskians that are
relevant to present discussion.

\section{The Schr\"{o}dinger equation}

\label{sec:Schro}

Before solving the Schr\"{o}dinger equation it is a good practice to convert
it into a dimensionless eigenvalue equation. In this way one removes all the
physical constants and reduces the number of model parameters to a minimum.
The time--independent Schr\"{o}dinger equation for a particle of mass $m$
that moves in one dimension ($-\infty <X<\infty $) under the effect of a
potential $V(X)$ is
\begin{equation}
-\frac{\hbar ^{2}}{2m}\psi ^{\prime \prime }(X)+V(X)\psi (X)=E\psi (X)
\label{eq:Schrodinger}
\end{equation}
where a prime indicates derivative with respect to the coordinate. If we
define the dimensionless coordinate $x=X/L$, where $L$ is an appropriate
length scale, then we obtain the dimensionless eigenvalue equation
\begin{eqnarray}
&&-\frac{1}{2}\varphi ^{\prime \prime }(x)+v(x)\varphi (x)=\epsilon \varphi
(x)  \nonumber \\
&&\varphi (x)=\sqrt{L}\psi (Lx),\;v(x)=\frac{mL^{2}}{\hbar ^{2}}
V(Lx),\;\epsilon =\frac{mL^{2}E}{\hbar ^{2}}  \label{eq:Schro_dim}
\end{eqnarray}
The length unit $L$ that renders both $\epsilon $ and $v(x)$
dimensionless is arbitrary and we can choose it in such a way that
makes the Schr\"{o}dinger equation simpler. We will see some
examples in section~\ref {sec:examples}.

It is well known that a general solution to the second--order differential
equation (\ref{eq:Schro_dim}) can be written as a linear combination of two
linearly independent solutions. Here we write
\begin{equation}
\varphi (x)=A_{2}C(x)+B_{2}S(x)  \label{eq:phi_C_S}
\end{equation}
where the solutions $C(x)$ and $S(x)$ satisfy
\begin{equation}
C(x_{0})=S^{\prime }(x_{0})=1,\;C^{\prime }(x_{0})=S(x_{0})=0
\label{eq:C,S_x0}
\end{equation}
at a given point $x_{0}$ in $(-\infty ,\infty )$. These conditions are
sufficient to ensure that $C(x)$ and $S(x)$ are linearly independent\cite
{PC61}.

For every value of the dimensionless energy $\epsilon $ we know that
\begin{equation}
\varphi (x)\rightarrow \left\{
\begin{array}{c}
A_{1}L_{c}(x)+B_{1}L_{d}(x),\,x\rightarrow -\infty \\
A_{3}R_{c}(x)+B_{3}R_{d}(x),\,x\rightarrow \infty
\end{array}
\right.  \label{eq:phi_asymp_gen}
\end{equation}
where $L$ and $R$ stand for left and right and $c$ and $d$ for
convergent and divergent, respectively. It means that, for
arbitrary $\epsilon $, the wavefunction is a linear combination of
a convergent and a divergent
function when $|x|\rightarrow \infty $. If, for a particular value of $%
\epsilon $, $B_{1}=B_{3}=0$ then the resulting wavefunction is square
integrable. In other words, this condition determines the energies of the
discrete spectrum.

It follows from the Wronskian properties outlined in the Appendix that
\begin{eqnarray}
B_{1}W(L_{c},L_{d}) &=&A_{2}W(L_{c},C)+B_{2}W(L_{c},S),\;x\rightarrow -\infty
\nonumber \\
B_{3}W(R_{c},R_{d}) &=&A_{3}W(R_{c},C)+B_{2}W(R_{c},S),\;x\rightarrow \infty
\end{eqnarray}
Therefore, when $B_{1}=B_{3}=0$ we have a linear homogeneous system of two
equations with two unknowns: $A_{2}$ and $B_{2}$. There will be nontrivial
solutions provided that its determinant vanishes
\begin{equation}
W(L_{c},C)W(R_{c},S)-W(R_{c},C)W(L_{c},S)=0  \label{eq:quant_cond_gen}
\end{equation}
The roots of this equation $\epsilon _{n}$, $n=0,1,\ldots $ are the energies
of the bound states (discrete spectrum).

When the potential is parity invariant
\begin{equation}
v(-x)=v(x)  \label{eq:v(x)_symm}
\end{equation}
and $x_{0}=0$ then $C(x)$ and $S(x)$ are even and odd functions,
respectively. In this case we have
\begin{eqnarray}
W(L_{c},S) &=&W(R_{c},S)  \nonumber \\
W(L_{c},C) &=&-W(R_{c},C)  \label{eq:W_symm}
\end{eqnarray}
and the determinant (\ref{eq:quant_cond_gen}) takes a simpler
form: $ W(R_{c},C)W(R_{c},S)=0$. We appreciate that the even and
odd solutions are clearly separate and their eigenvalues are given
by
\begin{eqnarray}
W(R_{c},C) &=&0  \nonumber \\
W(R_{c},S) &=&0  \label{eq:quant_cond_even_odd}
\end{eqnarray}
respectively. Besides, we need to consider only the interval $0\leq x<\infty
$.

Commonly, it is not difficult to derive approximate expressions
for the convergent and divergent asymptotic forms of the
wavefunction because they are straightforwardly determined by the
asymptotic behavior of the potential $v(x)$. Therefore, it only
remains to have sufficiently accurate expressions for $C(x)$ and
$S(x)$ and their derivatives in order to obtain the eigenvalues by
means of equation~(\ref{eq:quant_cond_gen}). This problem is
easily solved by means of, for example, a suitable numerical
integration method\cite{PFTV86}. If $y(x)$ stands for either
$C(x)$ or $S(x)$ then such an approach gives us its values at a
set of points $x_{0}-N_{L}h,
\,x_{0}-N_{L}h+h,\ldots ,\,x_{0},\,x_{0}+h,\ldots ,\,x_{0}+N_{R}h$ where $%
N_{L}$ and $N_{R}$ are the number of steps of size $h$ to the left
and right of $x_{0}$, respectively. The number of steps should be
sufficiently large so that $y(x)$ reaches its asymptotic value at
both $ x_{L}=x_{0}-N_{L}h $ and $x_{R}=x_{0}+N_{R}h$ and $h $
should be sufficiently small to provide a good representation of
$y(x)$. The numerical integration methods also yield the
derivative of the function $ y^{\prime }(x)$ at the same set of
points which facilitates the calculation of the Wronskians.

\section{Examples}

\label{sec:examples}

In order to test the accuracy of the Wronskian method we first
choose the exactly solvable problem given by the potential
$V(X)=-V_{0}/\cosh ^{2}(\alpha X)$, where $V_0 >0$ and $\alpha
>0$. If we set $L=1/\alpha $ we are led to the dimensionless
Schr\"{o}dinger equation (\ref{eq:Schro_dim}) where
\begin{eqnarray}
v(x) &=&-\frac{v_{0}}{\cosh ^{2}(x)},\;  \nonumber \\
v_{0} &=&\frac{mV_{0}}{\hbar ^{2}\alpha^2 },\;\epsilon =\frac{mE}{\hbar
^{2}\alpha^2 }  \label{eq:v(x)_cosh}
\end{eqnarray}
Note that the dimensionless energy $\epsilon$ depends on only one
independent potential parameter $v_{0}$. The units of length and energy are $%
1/\alpha $ and $\hbar ^{2}\alpha^2 /m$, respectively, and we do not have to
bother about the mass of the particle and the Planck constant when solving
the differential equation. The allowed dimensionless energies are given by%
\cite{F99}
\begin{eqnarray}
\epsilon _{n} &=&-\frac{1}{2}(\lambda -1-n)^{2},\;n=0,1,\ldots \leq \lambda
-1  \nonumber \\
\lambda &=&\frac{1}{2}\left( 1+\sqrt{1+8v_{0}}\right)  \label{eq:energ_exact}
\end{eqnarray}
and the spectrum is continuous for all $\epsilon >0$. It is clear that $%
\lambda \rightarrow 1$ as $v_{0}\rightarrow 0$ and there is only one bound
state when $1<\lambda <2$ ($0<v_{0}<1$). As $v_{0}$ increases more bound
states appear. As a result there are critical values of the potential
parameter for which $\epsilon _{n}=0$ that are given by the condition $%
\lambda _{n}=n+1$ or $v_{0,n}=\lambda _{n}(\lambda _{n}-1)/2=n(n+1)/2$.

Since $\lim_{|x|\rightarrow \infty }v(x)=0$ then in this case $%
R_{c}(x)=e^{-kx}$ and $R_{d}(x)=e^{kx}$, where $k^{2}=-2\epsilon $ (we only
consider the interval $0\leq x<\infty $ because the potential is parity
invariant). Consequently, the allowed energies are determined by the
conditions
\begin{eqnarray}
W(R_{c},C) &=&\left[ C^{\prime }(x)+kC(x)\right] e^{-kx}=0,\;x\rightarrow
\infty  \nonumber \\
W(R_{c},S) &=&\left[ S^{\prime }(x)+kS(x)\right] e^{-kx}=0,\;x\rightarrow
\infty  \label{eq:quant_cond_even_odd_2}
\end{eqnarray}
for even and odd states, respectively.

\begin{figure}[tbp]
\begin{center}
\par
\includegraphics[width=9cm]{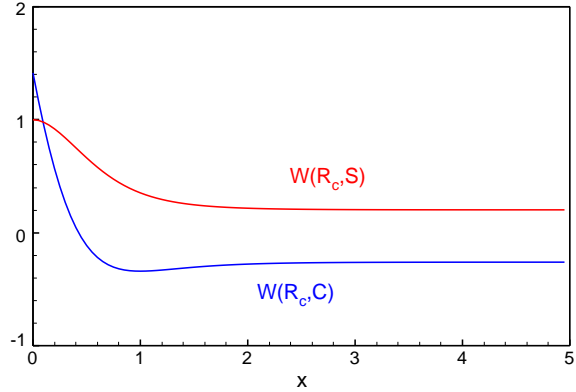}
\par
\end{center}
\caption{Wronskians for the exactly solvable problem with $\epsilon=-1$ and $%
v_0=2.5$} \label{fig:W_x}
\end{figure}

Since the potential (\ref{eq:v(x)_cosh}) is parity invariant we
integrate the Schr\"{o}dinger equation from $x_{0}=0$ to
$x_{R}=N_{R}h$. Fig.~\ref {fig:W_x} shows the Wronskians
$W(R_{c},C)$ and $W(R_{c},S)$ for $\epsilon =-1$ and $v_{0}=2.5$.
We appreciate that $x=x_{R}=5$ is large enough to have constant
asymptotic Wronskians and we choose this coordinate value from now
on. This numerical test also shows that it is sufficient for
present purposes to set $h=0.01$ and $N_{R}=500$ in the
fourth--order Runge--Kutta method\cite{PFTV86} built in the
computer algebra system Derive (http://www.chartwellyorke.
com/derive.html).

\begin{figure}[tbp]
\begin{center}
\par
\includegraphics[width=9cm]{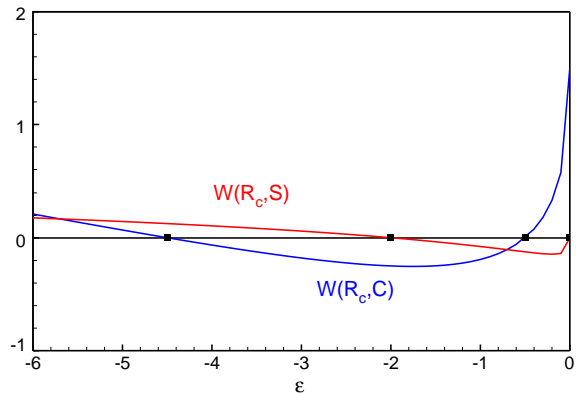}
\par
\end{center}
\caption{Wronskians for the exactly solvable problem with $v_0=6$
as functions of $\epsilon$. Squares mark the exact eigenvalues }
\label{fig:BS_ex}
\end{figure}

Fig.~\ref{fig:BS_ex} shows both $W(R_{c},C)$ and $W(R_{c},S)$ for
$x=5$ and $ v_{0}=6$ as functions of $\epsilon $. We see that the
Wronskians vanish at the exact eigenvalues given by equation
(\ref{eq:energ_exact}). This is a confirmation of our earlier
assumption that the number of steps and their size are suitable
for obtaining reasonable results.

\begin{figure}[tbp]
\begin{center}
\par
\includegraphics[width=9cm]{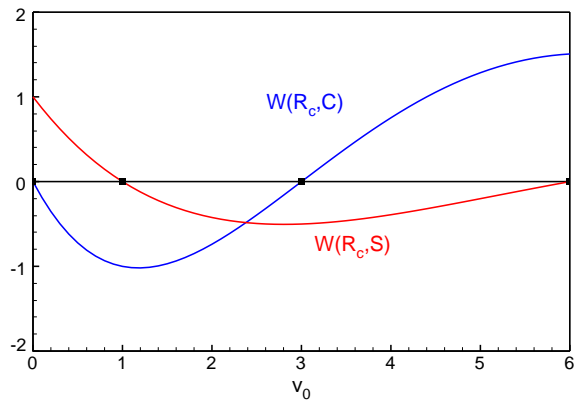}
\par
\end{center}
\caption{Wronskians for the exactly solvable problem with
$\epsilon=0$ as functions of $v_0$. Squares mark the exact
critical parameters } \label{fig:Crit_par}
\end{figure}

Fig.~\ref{fig:Crit_par} shows $W(R_{c},C)$ and $W(R_{c},S)$ for $x=5$ and $%
\epsilon =0$ as functions of $v_{0}$. In this case the Wronskians vanish at
the exact critical values $v_{0,n}=n(n+1)/2$.

\begin{figure}[tbp]
\begin{center}
\par
\includegraphics[width=9cm]{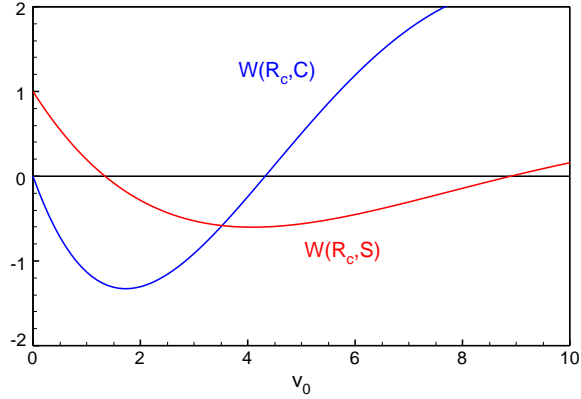}
\par
\end{center}
\caption{Wronskians for the Gaussian well with $\epsilon=0$ as functions of $%
v_0$ } \label{fig:Cr_pa_G}
\end{figure}

\begin{figure}[tbp]
\begin{center}
\par
\includegraphics[width=9cm]{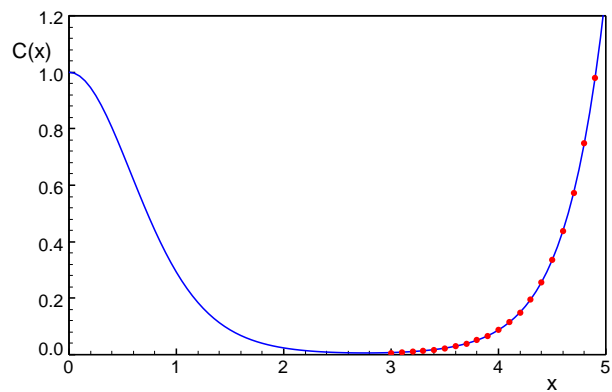}
\par
\end{center}
\caption{Unnormalized ground--state wavefunction for the Gaussian
well with $v_0=5$ } \label{fig:WFG}
\end{figure}

Those results for the exactly solvable problem suggest that the
Wronskian method is successful for all bound--state energies and
all well depths. As a second example, in what follows we try the
nontrivial problem provided by the Gaussian well
$V(x)=-V_{0}e^{-\alpha X^{2}}$, where $V_{0}>0$, and $\alpha >0$,
that we easily convert into the dimensionless potential
\begin{eqnarray}
v(x) &=&-v_{0}e^{-x^{2}},  \nonumber \\
v_{0} &=&\frac{mV_{0}}{\hbar ^{2}\alpha },\;\epsilon =\frac{mE}{\hbar
^{2}\alpha }  \label{eq:v(x)_Gaussian}
\end{eqnarray}
by means of the length unit $L=1/\sqrt{\alpha }$. This potential
is also parity invariant and vanishes asymptotically as
$|x|\rightarrow \infty $ so that the calculation is similar to the
preceding example. In order to show that the Wronskian method also
applies successfully to this model we first obtain some critical
values of the potential parameter $v_{0}$. Fig.~\ref {fig:Cr_pa_G}
shows $W(R_{c},C)$ and $W(R_{c},S)$ for $x=5$ and $\epsilon =0$ as
functions of $v_{0}$. The three zeros of the Wronskians shown in
the figure appear at $v_{0,1}=1.342$, $v_{0,2}=4.325$ and
$v_{0,3}=8.898$.
The numerical methods, although as accurate as one may desire (or
the computer allows) , are not exact; therefore we can make the
coefficient of the divergent part of the wavefunction very small
but never exactly equal to zero. To illustrate this point we
consider the ground state of the Gaussian well with $v_{0}=5$ and
obtain the approximate energy $\epsilon _{0}=-3.6077$ by means of
a naive bracketing algorithm. The Wronskian method gives us the
coefficient of the divergent part of the wavefunction
$B_{3}=W(R_{c},C)/W(R_{c},R_{d})=W(R_{c},C)/(2k)=1.886\times
10^{-6}$ as a by--product of the calculation at $x=5$. Note that
in this case $A_{2}=1$ and $B_{2}=0$ so that $C(x)$ is the
unnormalized approximation to $\varphi _{0}$. Fig.~\ref{fig:WFG}
shows that the wavefunction decreases as $ A_{3}e^{-kx}$ to almost
zero and then increases as $B_{3}e^{kx}$. For intermediate values
of the coordinate the convergent part dominates because $ A_{3}$
is much larger that $B_{3}$ but for larger values of $x$ $R_{d}$
dominates as shown in Fig.~\ref{fig:WFG}. The points in this
figure are the values of $B_{3}R_{d}=1.886\times
10^{-6}e^{2.686x}$.
We can improve the calculation and make $B_{3}$ as small as
desired, but it will never be exactly zero because the numerical
calculation is approximate and the spurious divergent part will
always be present. However, in order to normalize the wavefunction
and calculate expectation values we judiciously truncate the
coordinate interval at a convenient point and obtain finite
accurate values for those quantities. The Wronskian method is
clearly suitable for deciding on the truncation point and
estimating the error.

\section{Conclusions}

\label{sec:conclusions}

In our opinion the Wronskian method is sufficiently clear and
straightforward for teaching an advanced undergraduate or graduate course in
quantum mechanics. The mathematics requires no special background beyond an
introductory calculus course. Since many available computer softwares offer
numerical integration methods the programming effort is relatively light.

From a purely theoretical point of view the method is suitable for
the discussion of the convergent and divergent asymptotic
behaviors of the wavefunction and for illustrating how the allowed
bound--state energies make the divergent part vanish leading to
square--integrable wavefunctions. The students may try other
quantum--mechanical models, derive the appropriate asymptotic
behaviors analytically and then test their results by means of a
suitable computer program. They can verify that the Wronskians
already approach constants as the absolute value of the coordinate
increases and that the wavefunction already looks like a square
integrable function when the coefficient of the divergent
contribution is almost zero. They can even estimate the remnants
of the asymptotic divergent part because the Wronskian method
provides the necessary coefficients.

In addition to it, the Wronskian method is also suitable for quantum
scattering\cite{F11} allowing a unified treatment of both the discrete and
continuous spectra of the model.

Finally, we point out that computer algebra systems are remarkable aids for
the teaching and learning process because they facilitate the algebraic
treatment of the problem and even offer the possibility of straightforward
numerical calculations (although they are considerably slower than
specialized numerical programs).

\appendix

\section{Wronskians}

\label{sec:appendix}

In order to make this paper sufficiently self--contained in this appendix we
outline some well known results about the Wronskians that are useful for the
study of ordinary differential equations in general\cite{A69} and also for
the treatment of the Schr\"{o}dinger equation in particular\cite{PC61,WC73}.
To this end, we consider the ordinary second--order differential equation
\begin{equation}
L(y)=y^{\prime \prime }(x)+Q(x)y(x)=0  \label{eq:diffeq}
\end{equation}
If $y_{1}$ and $y_{2}$ are two linearly independent solutions to this
equation then we have
\begin{equation}
y_{1}L(y_{2})-y_{2}L(y_{1})=\frac{d}{dx}W(y_{1},y_{2})=0
\end{equation}
where
\begin{equation}
W(y_{1},y_{2})=y_{1}y_{2}^{\prime }-y_{2}y_{1}^{\prime }
\label{eq:Wronskian}
\end{equation}
is the Wronskian (or Wronskian determinant\cite{A69}). The Wronskian is a
skew--symmetric
\begin{equation}
W(f,g)=-W(g,f)\Rightarrow W(f,f)=0  \label{eq:Wronsk_prop_skew}
\end{equation}
and linear function of its arguments
\begin{eqnarray}
W(f_{1}+f_{2},g) &=&W(f_{1},g)+W(f_{2},g)  \nonumber \\
W(cf,g) &=&cW(f,g)  \label{eq:Wronsk_prop_lin}
\end{eqnarray}
where $c$ is a constant.

By linear combination of $y_{1}(x)$ and $y_{2}(x)$ we easily obtain two new
solutions $C(x)$ and $S(x)$ satisfying
\begin{equation}
C(x_{0})=S^{\prime }(x_{0})=1,\;C^{\prime }(x_{0})=S(x_{0})=0\;
\label{eq:C,S}
\end{equation}
at a given point $x_{0}$ so that $W(C,S)=1$ for all $x$. If we
write the general solution to equation~(\ref{eq:diffeq}) as
\begin{equation}
y(x)=AC(x)+BS(x)
\end{equation}
then
\begin{equation}
A=W(y,S),\;B=W(C,y)  \label{eq:A,B_W}
\end{equation}
This equation is quite useful for deriving relationships between the
coefficients of the asymptotic expansions of the wavefunction in different
regions of space as shown in sections \ref{sec:Schro} and \ref{sec:examples}%
. Additional mathematical properties of the Wronskians are available in
Powell and Crasemann's book on quantum mechanics\cite{PC61}.

\setcounter{section}{1}

\end{document}